\documentclass[aps,pra,10pt,amsmath,amssymb,onecolumn,notitlepage,tightenlines,floatfix,eqsecnum,showpacs]{revtex4-1}

\usepackage{graphicx}
\usepackage{subdepth}
\usepackage{mathtools}

\usepackage{xcolor}
\definecolor{dark-red}{rgb}{0.4,0.15,0.15}
\definecolor{dark-blue}{rgb}{0.15,0.15,0.4}
\definecolor{medium-blue}{rgb}{0,0,0.5}
\usepackage{hyperref}
\usepackage[all]{hypcap}
\usepackage{cleveref}
\hypersetup{
    colorlinks, linkcolor={dark-red},
    citecolor={dark-blue}, urlcolor={medium-blue}
} 

\renewcommand{\a}{a}

\newcommand{\ssp}{\hspace{0.4pt}}
\newcommand{\norm}[1]{\lvert #1 \rvert}
\newcommand{\normb}[1]{\big\lvert #1 \big\rvert}

\newcommand{\ket}[1]{\lvert\, #1\, \rangle}
\newcommand{\bra}[1]{\langle\, #1\, \rvert}
\newcommand{\braket}[2]{\langle\, #1\,\vert\, #2 \,\rangle}
\newcommand{\vac}{\mathrm{vac}}

\def\dbar{|\mkern-1.2mu|}
\def\mnorm#1{\dbar\mkern1.5mu#1\mkern0.1mu\dbar}

\begin{document}

\title{\bf Mixing nonclassical pure states in a linear-optical network almost always generates modal entanglement}

\date{\today}

\author{Zhang Jiang}
\affiliation{Center for Quantum Information and Control, University of New Mexico, MSC07-4220, Albuquerque, New Mexico 87131-0001, USA}

\author{Matthias D.~Lang}
\affiliation{Center for Quantum Information and Control, University of New Mexico, MSC07-4220, Albuquerque, New Mexico 87131-0001, USA}

\author{Carlton M.~Caves}
\affiliation{Center for Quantum Information and Control, University of New Mexico, MSC07-4220, Albuquerque, New Mexico 87131-0001, USA}

\begin{abstract}
In quantum optics a pure state is considered classical, relative to the statistics of photodetection, if and only if it is a coherent state.  A different and newer notion of nonclassicality is based on modal entanglement.  One example that relates these two notions is the Hong-Ou-Mandel effect, where modal entanglement is generated by a beamsplitter from the nonclassical photon-number state $\ket{1}\otimes\ket{1}$.  This suggests that beamsplitters or, more generally, linear-optical networks are mediators of the two notions of nonclassicality.  In this Brief Report, we show the following: Given a nonclassical pure-product-state input to an $N$-port linear-optical network, the output is almost always mode entangled; the only exception is a product of squeezed states, all with the same squeezing strength, input to a network that does not mix the squeezed and anti-squeezed quadratures.  Our work thus gives a necessary and sufficient condition for a linear network to generate modal entanglement from pure product inputs, a result that is of immediate relevance to the boson sampling problem.
\end{abstract}


\maketitle

\section{Introduction}
\label{sec:intro}

Beamsplitters are crucial elements in many applications, ranging from interferometry and homodyne detection to optical attenuation and quantum information and computation devices.  They are also essential components in many experiments designed to observe quantum effects, such as the Hong-Ou-Mandel effect~\cite{hong_measurement_1987}.  This effect, which underlies linear-optical quantum computation, occurs when two identical photons enter a 50:50 beamsplitter at the same time, one in each input port.  The two photons always exit the beamsplitter together in the same output mode, making the two output modes entangled.

Beamsplitters or, more generally, linear-optical networks are also crucial to the boson-sampling problem~\cite{aaronson_computational_2011, ralph_quantum_2013}, where the probability distribution of particular arrangements of bosons at the output of a linear-optical network is sampled.  It is an open question whether nonclassical input states other than Fock states lead to interesting sampling problems, i.e., problems cannot be efficiently simulated with a classical computer, and a necessary condition for an interesting problem is that the linear-optical network generates modal entanglement at the output.

In this Brief Report, we consider the relation between the nonclassicality of a pure-product-state input to an $N$-port linear-optical network (generalized beamsplitter) and the modal entanglement at the output of the network.  We use a notion of nonclassicality taken from quantum optics: the only classical pure states, relative to the statistics of photodetection, are the coherent states.  We demonstrate that given a nonclassical pure product-state input to an $N$-port linear-optical network, the output is almost always mode entangled; the only exception is a product of squeezed states, all with the same squeezing strength, input to a network that does not mix the squeezed and anti-squeezed quadratures.

A number of authors have discussed the relation between nonclassicality and entanglement of the input and output for a linear-optical network.  Kim~{\it et al}.~\cite{kim_entanglement_2002} conjectured that nonclassicality of the input state is a necessary condition for the network to act as a mode entangler, which was proved by Wang~\cite{xiang-bin_theorem_2002}.  Wolf~{\it et al}.~\cite{wolf_entangling_2003} established a connection, for Gaussian states, between squeezing at the input and the entanglement of the output.  Asb\'{o}th {\it et al}.~\cite{asboth_computable_2005} proposed to measure the nonclassicality of the input to an ordinary beamsplitter by the amount of entanglement that can be generated by the beamsplitter, ideal photodetectors, and auxiliary classical states.  Tahira~{\it et al}.~\cite{tahira_entanglement_2009} considered examples to link the single-mode nonclassicality of Gaussian states input to an ordinary beamsplitter to the robustness against noise of the entanglement at the output.

\section{Description of a linear-optical network}
\label{sec:beamsplitter}

An $N$-port device is described by a unitary operator $\mathcal{U}$ that takes the input state of $N$ modes to the output state,
\begin{align}
\ket{\Psi_{\rm out}}=\mathcal{U}\ssp \ket{\Psi_{\rm in}}\,.
\end{align}
A linear-optical network transforms the $N$ modal annihilation operators linearly among themselves, i.e., without mixing creation operators with annihilation operators.  The transformation can thus be written as
\begin{align}\label{eq:n_port_beamsplitter}
\a_j\rightarrow\mathcal{U}^\dagger \a_j\ssp\ssp\mathcal{U}
=\sum_{k=1}^N \a_k U_{kj}\,,
\end{align}
where $U$ is a matrix that must be unitary in order to preserve the canonical commutation relations.  By introducing a column vector of the annihilation operators, $\mathbf\a = \big(\a_1\;\a_2\;\cdots\;\a_N\big)^T$, we can
write the transformation~(\ref{eq:n_port_beamsplitter}) in the compact form
\begin{align}\label{eq:n_port_beamsplitter_two}
 \mathcal{U}^\dagger\mathbf\a\,\mathcal{U}=U^T \mathbf\a\,.
\end{align}
In terms of the column vector of creation operators, $\mathbf\a^\dagger = \big(\a_1^\dagger\;\a_2^\dagger\;\cdots\;\a_N^\dagger\big)^T$, the transformation is $\mathcal{U}^\dagger\mathbf\a^\dagger\mathcal{U}=U^\dagger\mathbf\a^\dagger$, which is equivalent to
\begin{align}\label{eq:n_port_beamsplitter_three}
\mathcal{U}\mathbf\a^\dagger\,\mathcal{U}^\dagger=U\mathbf\a^\dagger\,.
\end{align}
A linear-optical network preserves the number of photons.  We can multiply $U$ by a phase so that $\mathcal{U}\ssp \ket{\vac}=\ket{\vac}$; the matrix $U$ then stores all the information about $\mathcal{U}$.  Notice that multiplying $U$ by diagonal unitary matrices on the left and right amounts to rephasing the input and output modes without changing the physical transformation.

For the case of an ordinary beamsplitter, $N=2$, Eq.~(\ref{eq:n_port_beamsplitter_three}) has the general form (modulo irrelevant phase shifts),
\begin{align}
 \mathcal{U}
 \begin{pmatrix}
 \a_1^\dagger \\[4pt] \a_2^\dagger
 \end{pmatrix}
 \mathcal{U}^\dagger
 =
 \begin{pmatrix}
  \cos \theta                 &     e^{-i\varphi} \sin \theta \\[4pt]
  -e^{i\varphi} \sin \theta   &     \cos \theta
 \end{pmatrix}
 \begin{pmatrix}
  \a_1^\dagger \\[4pt] \a_2^\dagger
 \end{pmatrix}\,,
\end{align}
where $\theta \in [0, \pi/2]$ determines the reflectivity of the beamsplitter and $\varphi \in [0,2\pi)$ gives the phase difference of the output modes.  When $\theta = 0$ or $\pi/2$, the beamsplitter either does nothing or swaps the two modes; we exclude from our discussion these ``trivial'' beamsplitters, which merely permute the input modes.

Generally, for $N\ge2$, we impose the requirement that the linear-optical network cannot be decomposed into smaller networks acting on independent subsets of modes by permutations of the input and output modes, i.e., that we cannot turn $U$ into a direct sum of more than one block by a transformation $P_1 U P_2$, where $P_1$ and $P_2$ are permutation matrices.  We call such a network {\it connected}.  We lose no generality by considering only connected networks because in a disconnected network, each disconnected block can be studied separately.

One consequence of the connectedness is that no output mode is simply a rephased input mode.  Formally, this means that no row (column) of $U$ consists of zeros and a single element of magnitude one; stated differently, this {\it nontriviality}, which generalizes that for $N=2$, means that all elements of $U$ have magnitude strictly less than unity:
\begin{align}\label{eq:nontrivial_condition}
 \norm{U_{kj}} < 1\quad\mbox{for all $k,j\in 1,2,\ldots,N$}\,.
\end{align}

Now we can state precisely the question we address: What pure product states input to a connected $N$-port linear-optical network lead to a product output state?

\section{Bargmann-Fock representation}
\label{sec:BF}

The tool we use to answer this question is the Bargmann-Fock (BF) representation~\cite{bargmann_hilbert_1961, bargmann_remarks_1962, glauber_coherent_1963}, which maps every pure quantum state $\ket{\Psi}$ for $N$ modes to a particular analytic (holomorphic) function on $\mathbb{C}^N$,
\begin{align}
\begin{split}
B(\mathbf z)&=B(z_1,z_2,\dots,z_N)\\
&=e^{\norm{\mathbf{z}}^2/2}\,\braket{\overline{\mathbf{z}}}{\Psi}\\
&=\bigl\langle\,\vac\,\bigl|\,e^{{\mathbf{z}^T}\!\mathbf{a}}\bigr|\,\Psi\,\bigr\rangle\\
&=\sum_{n_1,\ldots,n_N}
\frac{\braket{n_1,\ldots,n_N}{\Psi}}{\sqrt{n_1!\cdots n_N!}}\,
z_1^{n_1}\cdots z_N^{n_N}\,.
\end{split}
\end{align}
Here an overbar denotes a complex conjugate, and
\begin{align}
\ket{\mathbf{z}}
=\mathcal{D}(\mathbf{z})\ket{\vac}
=e^{-\norm{\mathbf{z}}^2/2}e^{\mathbf{z}^T\!\mathbf{a}^\dagger}\ket{\vac}
\end{align}
is an $N$-mode coherent state, with
\begin{align}
\begin{split}
\mathcal{D}(\mathbf{z})
&=\prod_{j=1}^N e^{z_ja_j^\dagger-\overline z_ja_j}\\
&=\exp\bigl(\mathbf{z}^T\mathbf{a}^\dagger-\overline{\mathbf{z}}^T\mathbf{a}\bigr)
=e^{-\norm{\mathbf{z}}^2/2}e^{\mathbf{z}^T\!\mathbf{a}^\dagger}e^{-\overline{\mathbf{z}}^T\mathbf{a}}
\end{split}
\end{align}
being the multi-mode displacement operator.  The state is reconstructed from the BF representation by
\begin{align}
 \ket{\Psi} &= B\big(\mathbf\a^\dagger\big)\ket{\vac}\,.
\end{align}
The inner product of $\ket{\Psi}$ and $\ket{\Phi}=C(\mathbf\a^\dagger)\ket{\vac}$ is given in terms of the BF representation by
\begin{align}
\braket{\Phi}{\Psi}
=\int\frac{d^{2N}\!\mathbf{z}}{\pi^N}\,\overline C(\mathbf{z})B(\mathbf{z})e^{-\norm{\mathbf{z}}^2}\,.
\end{align}

A displaced state $\ket{\Phi}=\mathcal{D}(\mathbf{y})\ket{\Psi}$ has a BF representation
\begin{align}
C(\mathbf{z})
=\bigl\langle\,\vac\,\bigl\vert\,e^{{\mathbf{z}^T}\!\mathbf{a}}\,\mathcal{D}\bigl(\mathbf{y}\bigr)\vert\,\Psi\,\bigr\rangle
=B(\mathbf{z}-\overline{\mathbf{y}})e^{\mathbf{y}^T\!\mathbf{z}}e^{-\norm{\mathbf{y}}^2/2}\,.
\end{align}

The transformation~(\ref{eq:n_port_beamsplitter_two}) takes a very simple form in the BF representation,
 \begin{align}
 \begin{split}
 B_{\rm out}(\mathbf{z})
 &=\bigl\langle\,{\vac}\,\bigl\vert\,e^{{\mathbf{z}^T}\!\mathbf{a}}\,\mathcal{U}\bigr\vert\,\Psi_{\rm in}\bigr\rangle\\
 &=\bigl\langle\,{\vac}\,\bigl\vert\,e^{\mathbf{z}^T\!U^T\!\mathbf{a}}\bigr\vert\,\Psi_{\rm in}\bigr\rangle
 =B_{\rm in}(U\mathbf{z})\,.
 \label{eq:Btrans}
 \end{split}
 \end{align}
Hence, we also have
\begin{align}
\ket{\Psi_{\rm out}}=B_{\rm in}(U\mathbf\a^\dagger)\ket{\vac}\,.
\end{align}

For an $N$-mode pure-product input state,
\begin{align}
 \ket{\Psi_\mathrm{in}} = \ket{\psi_1}\otimes \ket{\psi_2}\otimes \cdots\otimes \ket{\psi_N}\;,
\end{align}
the BF representation is a product,
\begin{align}
 B_\mathrm{in}(\mathbf z) & = B_1\big(z_1\big)\, B_2\big(z_2\big) \cdots B_N\big(z_N\big)\,,
\end{align}
where $B_j\big(z_j\big)$ is the BF representation of the $j$th~mode.

We introduce the logarithm of the BF representation,
\begin{align}
 G(\mathbf z)=\ln B (\mathbf z)\;,
\end{align}
which is analytic (because the composition of two analytic functions is also analytic) everywhere in its domain except for singular points at the zeros of $B$.  For a product input state, we have
\begin{align}\label{eq:unentangled_condition}
G_\mathrm{in}(\mathbf z) &= G_1(z_1)+ G_2(z_2)+\cdots +G_N(z_N)\,.
\end{align}
Our question now reduces to the following: What states of the form $G_\mathrm{in}(\mathbf z)$, when input to a connected $N$-port linear-optical network, lead to an output,
\begin{align}
G_\mathrm{out}(\mathbf z) = G_\mathrm{in}(U \mathbf z),
\end{align}
that also separates into a sum over modes as in Eq.~(\ref{eq:unentangled_condition})?

Before moving on, notice that a displacement of the input state is equivalent to a displacement of the output state; i.e., since $\mathcal{U}\,\mathcal{D}(\mathbf{y})\,\mathcal{U}^\dagger = \mathcal{D}(U^T\!\mathbf{y})$,
\begin{align}
 \mathcal{U}\,\mathcal{D}(\mathbf{y})\ket{\Psi_\mathrm{in}}
 =\mathcal{D}(U^T\!\mathbf{y})\,\mathcal{U}\,\ket{\Psi_\mathrm{in}}\,.
\end{align}
The displacement operator is a product of local operators on the modes, so displacement does not change any modal entanglement properties of the input or the output.  In particular, displacement cannot change a product state into an entangled state.

\section{Conditions for a connected network not to generate entanglement}
\label{sec:not_entangled}

To address our main question, we use the Maclaurin expansion of $G_\mathrm{in}(\mathbf z)$ at the origin; as discussed above, this expansion requires that $\braket{\vac}{\Psi_{\rm in}}=B_\mathrm{in}(\mathbf 0)\neq 0$.  Although this does not generally hold, we can make it so, without changing the status of product states and entanglement, both at the input and output, by displacing the state so that it has a vacuum component, i.e., $\bra{\vac}\mathcal{D}(\mathbf{y})\ket{\Psi_\mathrm{in}}\neq0$.

This preliminary step having been taken, the Maclaurin expansion of $G_\mathrm{in}(\mathbf z)$ can be written as
\begin{align}
G_\mathrm{in}(\mathbf z)
&= \sum_{j=1}^N \bigg(\sum_{d=0}^\infty \lambda_j^{(d)} z_j^d\bigg)
=\sum_{d=0}^\infty\, \sum_{j=1}^N \lambda_j^{(d)} z_j^d\,,
\end{align}
where $\lambda_j^{(d)}$ is the $d$th expansion coefficient of $G_j(z)$. A product output state means that $G_\mathrm{out}(\mathbf z)$ has a similar expansion,
\begin{align}\label{eq:inout}
\sum_{d=0}^\infty\, \sum_{j=1}^N \xi_j^{(d)} z_j^d=
G_\mathrm{out}(\mathbf z)
=\sum_{d=0}^\infty\, \sum_{k=1}^N \lambda_k^{(d)} \bigg(\sum_{j=1}^N U_{kj}\ssp z_j\bigg)^{\!d} \,.
\end{align}
To save space below, we introduce column vectors for the input and output expansion coefficients at each $d$: $\boldsymbol{\xi}^{(d)}=\bigl(\xi_1^{(d)}\,\cdots\,\xi_N^{(d)}\bigr)^T$ and $\boldsymbol{\lambda}^{\!(d)}=\bigl(\lambda_1^{(d)}\,\cdots\,\lambda_N^{(d)}\bigr)^T$.

Analyticity requires that the equality in Eq.~(\ref{eq:inout}) hold order by order in $d$, i.e.,
\begin{align}\label{eq:unentangled_condition_dgen}
 \sum_{j=1}^N \xi_j^{(d)} z_j^d\,=\,\sum_{k=1}^N \lambda_k^{(d)}\bigg(\sum_{j=1}^N U_{kj}\ssp z_j\bigg)^{\!d} \,.
\end{align}
Indeed, analyticity further requires equality for each monomial in $z$; we now explore the consequences of this requirement.

For $d=0$, we have
\begin{align}
\sum_{j=1}^N\xi_j^{(0)}=\sum_{j=1}^N\lambda_j^{(0)}\,,
\end{align}
a condition that can always be satisfied.  For $d=1$, we have a straightforward linear relation between input and output,
\begin{align}
\xi_j^{(1)}=\sum_{k=1}^N\,\lambda_k^{(1)}U_{kj}
\quad\Longleftrightarrow\quad
\boldsymbol{\xi}^{(1)}=U^T\boldsymbol{\lambda}^{\!(1)}\,.
\end{align}

For $d = 2$, we have
\begin{align}
\xi_j^{(2)}\delta_{jj'}
= \sum_{k=1}^N \lambda_k^{(2)}\, U_{kj}\, U_{kj'}\,,
\end{align}
which can be rewritten in the more transparent form
\begin{align}\label{eq:unentangled_condition_2}
 \overline U_{kj}\,\xi_j^{(2)}= \lambda_k^{(2)}\,U_{kj}\;.
\end{align}
If $U$ is connected, Eq.~(\ref{eq:unentangled_condition_2}) can be satisfied only when $\norm{\xi^{(2)}_j} = \norm{\lambda^{(2)}_j} = {{\rm constant}}$ for all $j\in \{1,2,\ldots, N\}$.  Using the freedom to rephase input and output modes, we can make the coefficients $\xi_j^{(2)}$ and $\lambda_j^{(2)}$ real, and once this is done, Eq.~(\ref{eq:unentangled_condition_2}) implies that $U$ is real.

For $d=3$, the cross terms on the right-hand side of Eq.~(\ref{eq:unentangled_condition_dgen}), i.e., those monomials involving more than one $z_j$, must vanish, and for the non-cross terms, we must have
\begin{align}\label{eq:unentangled_condition_d}
  \xi_j^{(d)}
  &= \sum_{k=1}^N \lambda_k^{(d)} U_{kj}^d
  \quad\Longleftrightarrow\quad
  \boldsymbol{\xi}^{(d)}=U_d^T\boldsymbol{\lambda}^{\!(d)}
  \,,
\end{align}
where $U_d$ is the matrix whose elements are $d$th powers of the elements of $U$, i.e., $(U_d)_{jk}=U_{jk}^d$.  We now show that these conditions cannot be satisfied by any linear-optical network, unless $\boldsymbol{\lambda}^{\!(d)}$ vanishes.

Because a linear-optical network preserves photon number, the transformation~(\ref{eq:Btrans}) takes a monomial in $z_j$ to a polynomial of the same degree.  In particular, the transformation
\begin{align}
  \sum_{j=1}^N \lambda_j^{(d)}z_j^d
  \,\longrightarrow\,
  \sum_{k=1}^N \lambda_k^{(d)}\bigg(\sum_{j=1}^N U_{kj}\ssp z_j\bigg)^{\!d}
  =\sum_{j=1}^N \xi_j^{(d)} z_j^d
\end{align}
can be regarded as the transformation from the BF representation of an input state,
\begin{align}
\ket{\Phi_{\rm in}}=\sum_{j=1}^N \lambda_j^{(d)}(a_j^\dagger)^d\ket\vac\,,
\end{align}
which is normalized to
$\braket{\Phi_{\rm in}}{\Phi_{\rm in}}=d\ssp!\,\norm{\boldsymbol{\lambda}^{\!(d)}}^2$, to the BF representation of an output state, $\ket{\Phi_{\rm out}}=\mathcal{U}\ket{\Phi_{\rm in}}$, which is normalized to $\braket{\Phi_{\rm out}}{\Phi_{\rm out}}=d\ssp!\,\norm{\boldsymbol{\xi}^{(d)}}^2$.  Since the transformation preserves normalization, we must have
\begin{align}\label{eq:unentangled_condition_d2}
\normb{\boldsymbol{\lambda}^{\!(d)}}^2
=\normb{\boldsymbol{\xi}^{(d)}}^2
=\normb{U_d^T\ssp\boldsymbol{\lambda}^{\!(d)}}^2\,.
\end{align}

We now invoke several results about matrix norms (see Ref.~\cite{horn_matrix_1985}, esp.~Chap.~5 and Ex.~5.21) to write
\begin{align}
\frac{\normb{U_d^T\ssp\boldsymbol{\lambda}^{\!(d)}}^2}
{\norm{\boldsymbol{\lambda}^{\!(d)}}^2}
\le\mnorm{U_d^T}_2^2\le\mnorm{U_d}_1\,\mnorm{U_d^T}_1\,.
\end{align}
Here $\mnorm{M}_2$ is the spectral norm (2-norm) of $M$, i.e., the largest singular value of $M$ (square root of the largest eigenvalue of $M^\dagger M$), and $\mnorm M_1=\max_k\sum_j|M_{jk}|$ is the maximum-column-sum norm (1-norm) of $M$.  For $d\ge3$, the nontriviality condition~(\ref{eq:nontrivial_condition}) implies that the 1-norm of $U_d$ and $U_d^T$ are both strictly less than 1:{
\begin{align}
\mnorm{U_d}_1=\max_k\sum_j|U_{jk}^d|<\max_k\sum_j|U_{jk}|^2=1\,,
\end{align}
}
and likewise for $\mnorm{U_d^T}_1$.

Together these results mean that equality cannot be attained in Eq.~(\ref{eq:unentangled_condition_d2}) and thus that the output state cannot be a product state if $G_{\rm in}(\mathbf{z})$ has any terms of degree $d\ge3$.

Summarizing, we have the following circumstances where a pure-product input to a connected
linear-optical network produces a product-state output.

$\bullet$~~$G_{\rm in}(\mathbf z)$ has only constant and linear terms, i.e.,
\begin{align}\label{eq:Gin1}
G_\mathrm{in}(\mathbf z)
=-\frac12\norm{\boldsymbol\lambda}^2 + \boldsymbol\lambda^T\mathbf z\,.
\end{align}
This is the input coherent state
$e^{-\norm{\boldsymbol\lambda}^2/2}e^{\boldsymbol\lambda^T\mathbf{a}^\dagger}\ket\vac
=\ket{\boldsymbol\lambda}$.

$\bullet$~~$G_\mathrm{in}(\mathbf z)$ has quadratic, but not higher terms, and the input and output modes can be rephased so that
\begin{align}\label{eq:Gin2}
G_\mathrm{in}(\mathbf z)
=-\frac{N}{2}\ln(\cosh\gamma)-\frac12\boldsymbol\lambda^T\overline{\mathbf{y}}
+\boldsymbol\lambda^T\mathbf z-\frac{\tanh\gamma}{2}\,\mathbf z^T\mathbf z
\end{align}
and $U=\overline U$ is a real matrix.  Here the second-order coefficients are $\lambda_j^{(2)}=-\frac12\tanh\gamma$, $j=1,\ldots,N$, with $\gamma$ real, and $\mathbf y$ is written in terms of the real and imaginary parts of $\boldsymbol\lambda$ as $\mathbf y=(\boldsymbol\lambda_Re^{-\gamma}+i\boldsymbol\lambda_Ie^\gamma)\cosh\gamma$.  The input state is obtained by squeezing and then displacing the vacuum,
\begin{align}\label{eq:displaced_squeezed_vacuum}
\ket{\Psi_{\rm in}}
&=\frac{e^{-\boldsymbol\lambda^T\overline{\mathbf{y}}/2}
e^{\boldsymbol\lambda^T\mathbf{a}^\dagger}}{(\cosh\gamma)^{N/2}}
\prod_{j=1}^N\exp\biggl(-\frac{\tanh\gamma}{2}\, a_j^{\dagger\,2}\biggr)\ket\vac\nonumber\\
&=\mathcal{D}(\mathbf y)S_1(\gamma)\cdots S_N(\gamma)\ket\vac\,,
\end{align}
where
\begin{align}
 \mathcal{S}_j(\gamma)=e^{\gamma(\a_j^2-\a_j^{\dagger\,2})/2}
\end{align}
is the squeeze operator for the $j$th mode.  The imaginary part of $G_{\rm in}(\mathbf 0)$ in Eq.~(\ref{eq:Gin2}), which gives the overall phase of $\ket{\Psi_{\rm in}}$, is chosen so that the second equality in Eq.~(\ref{eq:displaced_squeezed_vacuum}) holds without an additional phase factor.

\section{Conclusion}
\label{sec:conclusion}
For coherent input states, no entanglement can be generated by any linear-optical network.  All nonclassical pure-product inputs to a connected network produce entanglement at the output, except particular kinds of displaced-squeezed vacuum states input to a particular sort of network.  The exception can always be reduced, by rephasing of input and output modes, to the situation where the input modes are all squeezed by the same amount along the same phase-space axes and the optical network acts identically on the squeezed and anti-squeezed quadratures without mixing them.

\begin{acknowledgments}
ZJ thanks S.~Pandey and S.~Rahimi-Keshari for useful discussions. This work was supported by National Science Foundation Grant Nos.~PHY-1212445 and PHY-1005540.
\end{acknowledgments}


\begin{thebibliography}{12}%
\makeatletter
\providecommand \@ifxundefined [1]{%
 \@ifx{#1\undefined}
}%
\providecommand \@ifnum [1]{%
 \ifnum #1\expandafter \@firstoftwo
 \else \expandafter \@secondoftwo
 \fi
}%
\providecommand \@ifx [1]{%
 \ifx #1\expandafter \@firstoftwo
 \else \expandafter \@secondoftwo
 \fi
}%
\providecommand \natexlab [1]{#1}%
\providecommand \enquote  [1]{``#1''}%
\providecommand \bibnamefont  [1]{#1}%
\providecommand \bibfnamefont [1]{#1}%
\providecommand \citenamefont [1]{#1}%
\providecommand \href@noop [0]{\@secondoftwo}%
\providecommand \href [0]{\begingroup \@sanitize@url \@href}%
\providecommand \@href[1]{\@@startlink{#1}\@@href}%
\providecommand \@@href[1]{\endgroup#1\@@endlink}%
\providecommand \@sanitize@url [0]{\catcode `\\12\catcode `\$12\catcode
  `\&12\catcode `\#12\catcode `\^12\catcode `\_12\catcode `\%12\relax}%
\providecommand \@@startlink[1]{}%
\providecommand \@@endlink[0]{}%
\providecommand \url  [0]{\begingroup\@sanitize@url \@url }%
\providecommand \@url [1]{\endgroup\@href {#1}{\urlprefix }}%
\providecommand \urlprefix  [0]{URL }%
\providecommand \Eprint [0]{\href }%
\providecommand \doibase [0]{http://dx.doi.org/}%
\providecommand \selectlanguage [0]{\@gobble}%
\providecommand \bibinfo  [0]{\@secondoftwo}%
\providecommand \bibfield  [0]{\@secondoftwo}%
\providecommand \translation [1]{[#1]}%
\providecommand \BibitemOpen [0]{}%
\providecommand \bibitemStop [0]{}%
\providecommand \bibitemNoStop [0]{.\EOS\space}%
\providecommand \EOS [0]{\spacefactor3000\relax}%
\providecommand \BibitemShut  [1]{\csname bibitem#1\endcsname}%
\let\auto@bib@innerbib\@empty
\bibitem [{\citenamefont {Hong}\ \emph {et~al.}(1987)\citenamefont {Hong},
  \citenamefont {Ou},\ and\ \citenamefont {Mandel}}]{hong_measurement_1987}%
  \BibitemOpen
  \bibfield  {author} {\bibinfo {author} {\bibfnamefont {C.~K.}\ \bibnamefont
  {Hong}}, \bibinfo {author} {\bibfnamefont {Z.~Y.}\ \bibnamefont {Ou}}, \ and\
  \bibinfo {author} {\bibfnamefont {L.}~\bibnamefont {Mandel}},\ }\bibfield
  {title} {\enquote {\bibinfo {title} {Measurement of subpicosecond time
  intervals between two photons by interference},}\ }\href {\doibase
  10.1103/PhysRevLett.59.2044} {\bibfield  {journal} {\bibinfo  {journal}
  {Phys. Rev. Lett.}\ }\textbf {\bibinfo {volume} {59}},\ \bibinfo {pages}
  {2044} (\bibinfo {year} {1987})}\BibitemShut {NoStop}%
\bibitem [{\citenamefont {Aaronson}\ and\ \citenamefont
  {Arkhipov}(2011)}]{aaronson_computational_2011}%
  \BibitemOpen
  \bibfield  {author} {\bibinfo {author} {\bibfnamefont {S.}~\bibnamefont
  {Aaronson}}\ and\ \bibinfo {author} {\bibfnamefont {A.}~\bibnamefont
  {Arkhipov}},\ }\bibfield  {title} {\enquote {\bibinfo {title} {The
  computational complexity of linear optics},}\ }in\ \href {\doibase
  10.1145/1993636.1993682} {\emph {\bibinfo {booktitle} {Proceedings of the
  43rd Annual {ACM} Symposium on Theory of Computing}}}\ (\bibinfo  {publisher}
  {{ACM}},\ \bibinfo {year} {2011})\ pp.\ \bibinfo {pages}
  {333--342}\BibitemShut {NoStop}%
\bibitem [{\citenamefont {Ralph}(2013)}]{ralph_quantum_2013}%
  \BibitemOpen
  \bibfield  {author} {\bibinfo {author} {\bibfnamefont {T.~C.}\ \bibnamefont
  {Ralph}},\ }\bibfield  {title} {\enquote {\bibinfo {title} {Quantum
  computation: Boson sampling on a chip},}\ }\href {\doibase
  10.1038/nphoton.2013.175} {\bibfield  {journal} {\bibinfo  {journal} {Nature
  Phot.}\ }\textbf {\bibinfo {volume} {7}},\ \bibinfo {pages} {514} (\bibinfo
  {year} {2013})}\BibitemShut {NoStop}%
\bibitem [{\citenamefont {Kim}\ \emph {et~al.}(2002)\citenamefont {Kim},
  \citenamefont {Son}, \citenamefont {Bu\v{z}ek},\ and\ \citenamefont
  {Knight}}]{kim_entanglement_2002}%
  \BibitemOpen
  \bibfield  {author} {\bibinfo {author} {\bibfnamefont {M.~S.}\ \bibnamefont
  {Kim}}, \bibinfo {author} {\bibfnamefont {W.}~\bibnamefont {Son}}, \bibinfo
  {author} {\bibfnamefont {V.}~\bibnamefont {Bu\v{z}ek}}, \ and\ \bibinfo
  {author} {\bibfnamefont {P.~L.}\ \bibnamefont {Knight}},\ }\bibfield  {title}
  {\enquote {\bibinfo {title} {Entanglement by a beam splitter: Nonclassicality
  as a prerequisite for entanglement},}\ }\href {\doibase
  10.1103/PhysRevA.65.032323} {\bibfield  {journal} {\bibinfo  {journal} {Phys.
  Rev.~A}\ }\textbf {\bibinfo {volume} {65}},\ \bibinfo {pages} {032323}
  (\bibinfo {year} {2002})}\BibitemShut {NoStop}%
\bibitem [{\citenamefont {Wang}(2002)}]{xiang-bin_theorem_2002}%
  \BibitemOpen
  \bibfield  {author} {\bibinfo {author} {\bibfnamefont {X.-B.}\ \bibnamefont
  {Wang}},\ }\bibfield  {title} {\enquote {\bibinfo {title} {Theorem for the
  beam-splitter entangler},}\ }\href {\doibase 10.1103/PhysRevA.66.024303}
  {\bibfield  {journal} {\bibinfo  {journal} {Phys.\ Rev.~A}\ }\textbf
  {\bibinfo {volume} {66}},\ \bibinfo {pages} {024303} (\bibinfo {year}
  {2002})}\BibitemShut {NoStop}%
\bibitem [{\citenamefont {Wolf}\ \emph {et~al.}(2003)\citenamefont {Wolf},
  \citenamefont {Eisert},\ and\ \citenamefont {Plenio}}]{wolf_entangling_2003}%
  \BibitemOpen
  \bibfield  {author} {\bibinfo {author} {\bibfnamefont {M.~M.}\ \bibnamefont
  {Wolf}}, \bibinfo {author} {\bibfnamefont {J.}~\bibnamefont {Eisert}}, \ and\
  \bibinfo {author} {\bibfnamefont {M.~B.}\ \bibnamefont {Plenio}},\ }\bibfield
   {title} {\enquote {\bibinfo {title} {Entangling power of passive optical
  elements},}\ }\href {\doibase 10.1103/PhysRevLett.90.047904} {\bibfield
  {journal} {\bibinfo  {journal} {Phys. Rev. Lett.}\ }\textbf {\bibinfo
  {volume} {90}},\ \bibinfo {pages} {047904} (\bibinfo {year}
  {2003})}\BibitemShut {NoStop}%
\bibitem [{\citenamefont {Asb\'{o}th}\ \emph {et~al.}(2005)\citenamefont
  {Asb\'{o}th}, \citenamefont {Calsamiglia},\ and\ \citenamefont
  {Ritsch}}]{asboth_computable_2005}%
  \BibitemOpen
  \bibfield  {author} {\bibinfo {author} {\bibfnamefont {J.~K.}\ \bibnamefont
  {Asb\'{o}th}}, \bibinfo {author} {\bibfnamefont {J.}~\bibnamefont
  {Calsamiglia}}, \ and\ \bibinfo {author} {\bibfnamefont {H.}~\bibnamefont
  {Ritsch}},\ }\bibfield  {title} {\enquote {\bibinfo {title} {Computable
  measure of nonclassicality for light},}\ }\href {\doibase
  10.1103/PhysRevLett.94.173602} {\bibfield  {journal} {\bibinfo  {journal}
  {Phys. Rev. Lett.}\ }\textbf {\bibinfo {volume} {94}},\ \bibinfo {pages}
  {173602} (\bibinfo {year} {2005})}\BibitemShut {NoStop}%
\bibitem [{\citenamefont {Tahira}\ \emph {et~al.}(2009)\citenamefont {Tahira},
  \citenamefont {Ikram}, \citenamefont {Nha},\ and\ \citenamefont
  {Zubairy}}]{tahira_entanglement_2009}%
  \BibitemOpen
  \bibfield  {author} {\bibinfo {author} {\bibfnamefont {R.}~\bibnamefont
  {Tahira}}, \bibinfo {author} {\bibfnamefont {M.}~\bibnamefont {Ikram}},
  \bibinfo {author} {\bibfnamefont {H.}~\bibnamefont {Nha}}, \ and\ \bibinfo
  {author} {\bibfnamefont {M.~S.}\ \bibnamefont {Zubairy}},\ }\bibfield
  {title} {\enquote {\bibinfo {title} {Entanglement of {Gaussian} states using
  a beam splitter},}\ }\href {\doibase 10.1103/PhysRevA.79.023816} {\bibfield
  {journal} {\bibinfo  {journal} {Phys. Rev.~A}\ }\textbf {\bibinfo {volume}
  {79}},\ \bibinfo {pages} {023816} (\bibinfo {year} {2009})}\BibitemShut
  {NoStop}%
\bibitem [{\citenamefont {Bargmann}(1961)}]{bargmann_hilbert_1961}%
  \BibitemOpen
  \bibfield  {author} {\bibinfo {author} {\bibfnamefont {V.}~\bibnamefont
  {Bargmann}},\ }\bibfield  {title} {\enquote {\bibinfo {title} {On a {Hilbert}
  space of analytic functions and an associated integral transform},}\ }\href
  {\doibase 10.1002/cpa.3160140303} {\bibfield  {journal} {\bibinfo  {journal}
  {Commun. Pure Appl. Math.}\ }\textbf {\bibinfo {volume} {14}},\ \bibinfo
  {pages} {187–214} (\bibinfo {year} {1961})}\BibitemShut {NoStop}%
\bibitem [{\citenamefont {Bargmann}(1962)}]{bargmann_remarks_1962}%
  \BibitemOpen
  \bibfield  {author} {\bibinfo {author} {\bibfnamefont {V.}~\bibnamefont
  {Bargmann}},\ }\bibfield  {title} {\enquote {\bibinfo {title} {Remarks on a
  {Hilbert} space of analytic functions},}\ }\href
  {http://www.ncbi.nlm.nih.gov/pmc/articles/PMC220756/} {\bibfield  {journal}
  {\bibinfo  {journal} {Proc. Natl. Acad. Sci. USA}\ }\textbf {\bibinfo
  {volume} {48}},\ \bibinfo {pages} {199} (\bibinfo {year} {1962})}\BibitemShut
  {NoStop}%
\bibitem [{\citenamefont {Glauber}(1963)}]{glauber_coherent_1963}%
  \BibitemOpen
  \bibfield  {author} {\bibinfo {author} {\bibfnamefont {R.~J.}\ \bibnamefont
  {Glauber}},\ }\bibfield  {title} {\enquote {\bibinfo {title} {Coherent and
  incoherent states of the radiation field},}\ }\href {\doibase
  10.1103/PhysRev.131.2766} {\bibfield  {journal} {\bibinfo  {journal} {Phys.
  Rev.}\ }\textbf {\bibinfo {volume} {131}},\ \bibinfo {pages} {2766} (\bibinfo
  {year} {1963})}\BibitemShut {NoStop}%
\bibitem [{\citenamefont {Horn}\ and\ \citenamefont
  {Johnson}(1985)}]{horn_matrix_1985}%
  \BibitemOpen
  \bibfield  {author} {\bibinfo {author} {\bibfnamefont {R.~A.}\ \bibnamefont
  {Horn}}\ and\ \bibinfo {author} {\bibfnamefont {C.~R.}\ \bibnamefont
  {Johnson}},\ }\href@noop {} {\emph {\bibinfo {title} {Matrix Analysis}}}\
  (\bibinfo  {publisher} {Cambridge University Press},\ \bibinfo {address}
  {Cambridge, England},\ \bibinfo {year} {1985})\BibitemShut {NoStop}%
\end{thebibliography}

%

\end{document}